\magnification=\magstep1
\vsize=9.5 truein
\hsize=6.5 truein
\hoffset=-0.1truein
\footline={\ifnum\pageno>1 \hss\tenrm --\folio-- \hss 
\else\hfill\fi} \maxdepth=2pt
\font\Atitle=cmr10 scaled\magstep2
\font\text=cmr12 at 12truept
\baselineskip=20pt

\text
\centerline{}
\vfill

\centerline{\Atitle Regimes of Precursor-Mediated Epitaxial 
Growth} \bigskip\bigskip\bigskip

\centerline{A. Zangwill}
\centerline{\it School of Physics, Georgia Institute of Technology, 
Atlanta,
Georgia 30332}
\bigskip

\centerline{D.D. Vvedensky}
\centerline{\it The Blackett Laboratory, Imperial College, London 
SW7 2BZ, United
Kingdom}
\bigskip\bigskip

\centerline{\bf Abstract}

\noindent
A discussion of epitaxial growth is presented for those situations 
(OMVPE, CBE, ALE, MOMBE, GSMBE, etc.) when the kinetics of 
surface processes associated with
molecular precursors may be rate limiting. Emphasis is placed on 
the identification of various {\it characteristic length scales} 
associated with the surface processes. Study of the relative 
magnitudes of these lengths permits one to identify regimes of 
qualitatively different growth kinetics as a function of 
temperature and deposition flux.
The approach is illustrated with a simple model which takes 
account of deposition, diffusion, desorption, dissociation, and step 
incorporation of a single precursor species, as well as the usual 
processes of atomic diffusion and step incorporation. Experimental 
implications are discussed in some detail. 

\vfill
\eject

A well-founded
conceptual and computational
framework now exists for the theoretical description of molecular 
beam epitaxy (MBE).
Surprisingly, it has
proved sufficient to focus almost exclusively on the diffusion and 
incorporation kinetics of single adatoms both in atomistic Monte 
Carlo simulations$^{1,2}$ and in studies based on reaction-diffusion
equations.$^{3-5}$ Within this context, not only is there  an analytic theory of
the transition between step flow and two-dimensional (2D) island nucleation and
coalescence$^{4-6}$, as monitored, e.g.,  by the disappearance of reflection
high-energy electron  diffraction (RHEED) oscillations
as a function of growth conditions but, for GaAs(001) under sufficiently 
As-rich conditions, {\it quantitative} agreement with RHEED 
oscillation data$^{2}$ can
be
achieved without any explicit reference to the As source (whether 
As$_2$ or As$_4$) in the theoretical model.  Unfortunately, there 
is increasing evidence that these simple adatom models are {\it 
not} adequate to describe the growth kinetics for the general case 
when the atomic constituents of the growing film are delivered to 
the substrate in the form of heteroatomic molecules. For example, the very different kinetics observed 
for surface reactions$^{7,8}$ and growth$^{9,10}$ when 
trimethylgallium (TMG) or
triethylgallium (TEG) is deposited onto GaAs makes clear that no 
single universal
theoretical model is likely to be proposed for this case.

The purpose of the present Letter is to demonstrate the existence 
and nature of
various qualitatively distinct kinetic regimes of epitaxial growth 
which can occur
whenever
{\it surface} chemical processes are
important.
We are motivated by detailed surface diffraction studies of GSMBE 
(using electrons)$^{11-13}$ and OMVPE (using x-rays)$^{14,15}$ 
which yield considerable information about the morphological 
evolution of the surface as growth proceeds. Since the
existence and evolution of {\it characteristic length scales} emerge 
quite naturally from such studies we organize our discussion 
around the identification
of these quantities.
For this purpose, it is essential to take explicit account of surface 
diffusion, a feature missing from essentially all existing theoretical 
models of such growth. The latter$^{16,17}$ typically take the 
form of
spatially uniform
coupled rate
equations and focus attention on the prediction of quantities such 
as the net growth rate and the average concentration of various 
surface species.
Nonetheless, as will become clear below, these earlier studies 
readily can be generalized so that the rather different point of
view we advocate here can be adopted.

To illustrate
our method, we first describe a simple growth scenario and then 
work out the qualitative
consequences which might be observed in a diffraction 
experiment of the sort noted earlier. We regard the model as a 
physically reasonable {\it minimal} generalization of the adatom 
models used to study MBE. It
is not intended to describe
any particular material system in detail. 

Consider deposition onto a vicinal surface of a molecule that 
contains the atomic constituent of the growing substrate. We 
include (see Fig. 1) the
processes of (i) desorption of the molecule back into the gas phase, 
(ii)
surface diffusion of the molecule in a weakly bound precursor 
state, (iii)
decomposition of the molecule on a terrace to release the atomic 
constituent, and (iv) decomposition of the molecule at a step edge 
and incorporation there of the atomic species. Included as well are 
(v) surface diffusion of the atomic species and (vi) incorporation 
of these atoms at a step edge. For
simplicity only, we
ignore all site-blocking effects and the 
presence (and fate) of all molecular decomposition fragments.  Desorption of the
atomic species is also neglected.

A quantitative theory results if we generalize the analysis of 
Burton, Cabrera and Frank$^{18}$ (BCF)
to include the processes described above. The one-dimensional continuum
model of BCF often is used to analyze MBE on vicinal surfaces.$^{3}$ In 
the present
case, the
evolution in time and space of the surface concentration of 
precursor molecules $n(x,t)$ and adatoms $c(x,t)$ is determined 
by the following coupled reaction-diffusion equations:
$$
\displaylines{
\hfill{\partial n\over\partial t}=D_M{\partial^2 n \over \partial 
x^2 } - {n\over\tau}-\kappa n+J\hfill(1)\cr
\noalign{\vskip3pt}
\hfill{\partial c\over\partial t}=D_A{\partial^2 c \over \partial 
x^2}+\kappa n\hfill(2)\cr}
$$
Here, $D_M$ and $D_A$ respectively denote the surface diffusion 
constant of the precursor molecule and the atom, $\kappa$ and 
$\tau^{-1}$
are the rate constants for decomposition and desorption of the 
precursor molecule on a terrace, and $J$ is the molecular 
deposition flux. Equations (1) and (2) are
supplemented by the boundary conditions
$$
D_A c_x(0,t)=\beta_A
[c(0,t)-c_0],\qquad - D_A c_x(
\ell,t)=\beta_A [c(\ell,t)-c_0] \eqno(3)
$$
$$
D_M n_x(0,t)=\beta_M n(0,t),\qquad -
D_M n_x(\ell,t)=\beta_M n(\ell,t)
\eqno(4)
$$
where $\ell$ is the terrace 
length and $c_0$ is the equilibrium
concentration of
atoms at the step edge. As discussed in detail elsewhere,$^{3}$ the 
choice (3)
indicates that atoms incident on a step incorporate into the solid 
at a rate proportional to $\beta_A$.
The choice
(4) guarantees that every precursor molecule incident on a step 
decomposes (and its atomic constituent incorporated into the step) 
at a rate proportional to $\beta_M$.  The steady-state (time-independent) solutions
of these equations  have simple analytic 
solutions$^{3}$ and one obtains an exact expression for the growth 
rate. We do not exhibit the explicit formula here since our aim in 
this paper is to identify the qualitative regimes of
growth implied by (1)--(4). Instead, we proceed to the 
identification of the relevant length scales. 

We begin with
$x_s=\sqrt{D_M \tau}$ and
$\ell_\kappa=\sqrt{D_M/\kappa}$. The quantity $x_s$ is the 
average distance a
molecule diffuses before desorbing while $\ell_\kappa$ is the 
average distance a
molecule diffuses before decomposing to release an adatom. Thus, 
the first two
regimes of importance are distinguished by whether 
$x_s/\ell_\kappa\gg1$ or
$x_s/\ell_\kappa\ll1$, i.e., whether the diffusing species are predominantly
adatoms or molecules. In the first case, we
must consider the pertinent length scales associated with the 
kinetics of the adatoms. The quantity $J_{\rm 
eff}(x)=\kappa n(x)$ is the effective
``flux'' of adatoms due to the decomposition of molecules. It then 
is natural to define the length $\ell_A=\root 4\of {D_A/ J_{\rm 
eff}}$, where $J_{\rm eff}$ is the
constant value obtained by averaging $J_{\rm eff}(x)$ over a 
terrace. Clearly,
$\ell_A$ corresponds to the distance an adatom diffuses before 
another adatom is released by a molecular decomposition reaction. 

There
now are two
new regimes to consider: $\ell_A/\ell\ll1$ and $\ell_A/\ell\gg1$. 
In the first
case, the typical migration distance of a free adatom is much 
smaller than the terrace
length, the
encounter probability of adatoms is high, and growth proceeds by 
the nucleation, growth and
coalescence of 2D islands on the terraces. In the 
second case, adatoms
diffuse to
the step edges before another atom is released, so growth 
proceeds by the advancement of steps, i.e., step flow. We note that 
the usefulness of the length scale $\ell_A$ has been demonstrated 
previously$^{5}$ for the case of MBE where $J_{\rm eff}$ is 
replaced by the vapor phase deposition flux of atoms. 

Within the atomic step flow
regime, there are two additional possibilities to consider. To see 
this, we define
a
length $d_A=D_A/\beta_A$ which is the {\it
additional}
distance an atom diffuses (after its
first arrival at a step)
before incorporation into the solid occurs. If $d_A \ll\ell$, 
incorporation
occurs
very soon after the first encounter with a step. We refer to this as
``fast atomic step flow.'' Clearly,
$\ell_A\gg\ell\gg d_A$ in this regime.
On the other hand, if
$d_A \gg\ell$, step edge incorporation requires several attempts. 
If $d_A \ll\ell_A$, growth
is reaction-limited at the step edges of the original surface and 
occurs by ``slow atomic step flow''. However, if $d_A\gg\ell_A$, 
attractive interactions between atoms on
the terraces become important and growth occurs by a 
combination of
slow step flow and 2D island formation. For growth 
under typical conditions of MBE, this latter mode appears to be 
appropriate.$^{19,20}$
In this regime we have $d_A \gg\ell_A\gg\ell$. 

If $x_s/\ell_\kappa\ll1$, the migrating species are predominantly 
the molecular precursors. To determine the growth regimes in this 
case, we first
construct the molecular analogue to $\ell_A$, i.e., $\ell_M=\root 
4\of {D_M/J}$.
The
quantity $\ell_M$ is the average distance a molecule travels 
before encountering another
molecule deposited by the incoming flux. We thus are lead to 
consider the regimes
$\ell_M/\ell\ll1$ and $\ell_M/\ell\gg1$. In the first case the 
molecules diffuse
to the step edge before encountering another molecule, while in 
the second case,
the molecules collide on the terraces before arriving at the step 
edge. Both
cases require further consideration.

If $\ell_M/\ell\ll1$, the encounter probability of the molecules on 
the
terraces is high and several scenarios can occur depending upon 
the precursor mean density. Possibilities include collisional 
decomposition with concomitant island nucleation, the
formation of a molecular film, and the formation of a liquid-like 
state of adsorbed molecules. Since our model does not include 
interactions among the
molecules we refer to this simply as the ``molecular interaction'' 
regime.

If $\ell_M/\ell\gg1$, it is useful to determine whether one or 
several encounters with a step are required before the molecule 
decomposes. This issue is
addressed by introducing the length $d_M=D_M/{\beta_M}$ 
which may be regarded
as the distance a precursor diffuses before a successful step-
catalyzed decomposition and incorporation reaction occurs. Again, 
two regimes are distinguished: $d_M/x_s\ll1$ or $d_M/x_s\gg1$. 
If $d_M/x_s\ll1$, most of the molecules immediately decompose 
upon arrival
at a step edge.
We
call this the regime of ``fast molecular step flow''. Note that 
$\ell_\kappa\gg x_s\gg d_M$ in this case. 

When $d_M/x_s\gg1$, we must specify the
relative magnitudes of $d_M$ and $\ell_\kappa$ since both are 
much greater than the mean desorption length $x_s$. If 
$\ell_\kappa$
is the shorter of the two, the molecule decomposes on the terrace 
and the growth mode is determined by atomic kinetics as 
discussed earlier. But if
$d_M$ is
the
shorter length, the molecules decompose predominantly at the 
step edge, albeit with some difficulty. We refer to this latter 
regime as ``slow molecular step flow'' to indicate that
even though molecular diffusion to the step edge is fast, the 
subsequent attachment of the atom is slow. 

The preceding discussion can be summarized conveniently in 
graphical form. To do so, note that the model parameters $D_A$, 
$D_M$, $\kappa$, $\tau$, $\beta_A$ and $\beta_M$ typically 
exhibit an Arrhenius form. Thus, the molecular deposition flux $J$, 
the mean terrace width $\ell$ and the substrate temperature $T$ 
are the true control parameters of the problem. We choose to 
display the various regimes discussed above as a ``kinetic phase 
diagram'' in the $T$-$J$ plane (Figure 2).
This diagram was generated by making several physically 
reasonable choices for the
energy barriers and pre-exponential
factors in the foregoing rate constants and computing the relative 
magnitudes of the various characteristic lengths.
Only the gross
topology of the phase fields should be noted since the precise 
placement of the phase boundaries depends on the details of the 
system in question.

What are the experimental implications of such a diagram? For 
definiteness, consider a diffraction
experiment where the time evolution of the intensities of both 
Bragg reflections and diffuse scattering are recorded. By tuning 
the control parameters so that oscillations in the former 
disappear, one straddles one of the curved phase boundaries in 
Figure 2 (where either $\ell=\ell_M$ or $\ell=\ell_A$) and the 
surface diffusion constant of the
primary migrating species can be extracted. By comparison with 
corresponding MBE results, one likely can determine if one is to 
the left or right of the long vertical line in Figure 2. At elevated 
deposition rates,
the diffuse scattering from such an experiment reveals the time 
evolution of the mean size of 2D islands and their 
average separation. An abrupt change in these quantities with 
increasing temperature could be
interpreted consistently (from Figure 2) as a kinetic transition 
from a regime of
molecular
interaction to a regime of 2D atomic island formation. Recent 
extensive
surface x-ray scattering
measurements of OMVPE growth of GaAs(001) onto a slightly 
misoriented substrate$^{14,15}$
provide an attractive data base for an analysis of the sort we 
propose.

As a more concrete example, consider a recent systematic study of 
the disappearance of RHEED oscillations as a function of 
temperature for GaAs(001) grown by MOMBE using TMG as the 
source of gallium.$^{10}$ A transition to step flow is observed to 
occur for both vicinal $A$-type surfaces (Ga-terminated step 
edges) and vicinal $B$-type surfaces
(As-terminated step edges). But, when compared to the result 
found in MBE, the transition temperature
$T_C$ is {\it increased} on the $A$ surface and {\it decreased} on 
the $B$ surface, both by approximately 30$\,^{\circ}$C. We 
speculate that this behavior reflects the presence of slow 
molecular step flow where
As-terminated
step sites are required to catalyze precursor$^{21}$ decomposition. 
In that way, the $B$ surface would be in slow step flow 
automatically. A reduced $T_C$ is expected due to the presence of 
Ga-terminated kink sites exposed by thermal fluctuations of the 
step edge. Conversely, more temperature is needed to effect step 
flow on the $A$ surface since a significant density of As-
terminated kink sites (again exposed due to thermal fluctuations) 
is required in that case. 

Evidently, the simple model described in this paper must be 
generalized if additional important kinetic steps are demonstrated 
to exist for any particular growth problem. This would complicate 
both the length scale analysis and the construction of the kinetic 
phase diagram. Nonetheless, such an exercise fosters a mode of 
thinking which, we believe, ultimately will help create a 
conceptual framework for the chemically-based epitaxial growth 
techniques comparable to that already achieved for MBE. 

The authors thank Ahmet Erbil,
Tom Foxon, Bruce Joyce, Tadaaki Kaneko and Tomoya Shitara for 
valuable discussions.
Work performed at Georgia Tech is supported by the U.S. 
Department of Energy under
grant No. DE-FG05-88ER45369. Work performed at Imperial 
College is supported by
Imperial College and the Research Development Corporation of 
Japan under the
auspices of the ``Atomic Arrangement: Design and Control for New 
Materials'' Joint
Research Program.
\vfill
\eject

{\frenchspacing
\centerline{\bf References}

\item{1.} A. Madhukar and S.V. Ghaisas, CRC Crit. Rev. Solid State 
Mater. Sci.
{\bf 13}, 1434 (1987).

\item{2.} T. Shitara, D.D. Vvedensky, M.R. Wilby, J. Zhang, J.H. Neave 
and B.A.
Joyce, Appl. Phys. Lett. {\bf 60}, 1504 (1992). 

\item{3.} R. Ghez and S.S. Iyer, IBM J. Res. Develop. {\bf 32}, 804 
(1988). 

\item{4.} A.K. Myers-Beaghton and D.D. Vvedensky, Phys. Rev. A 
{\bf 44}, 2457
(1991).

\item{5.} A.K. Myers-Beaghton and D.D. Vvedensky, Phys. Rev. B 
{\bf 42}, 5544 (1990).

\item{6.} V. Fuenzalida, Phys. Rev. B {\bf 44}, 10835 (1991). 

\item{7.} V.M. Donnelly, J.A. McCaulley and R.J. Shul, in {\it 
Chemical Perspectives of Microelectronic Materials II}, edited by 
L.V.Interrante, K.F.Jensen, L.H. Dubois and M.E. Gross (Materials 
Research Society, Pittsburgh, 1991), pp. 15-23.

\item{8.} M.L. Yu, U. Memmert, N.I. Buchan and T.F. Kuech, in {\it 
Chemical Perspectives of Microelectronic Materials II}, edited by 
L.V.Interrante, K.F.Jensen, L.H. Dubois and M.E. Gross (Materials 
Research Society, Pittsburgh, 1991), pp. 37-46.

\item{9.} Y. Okuno, H. Asahi, T. Kaneko, T.W. Kang, and S. Gonda, 
J. Crystal Growth {\bf 105}, 185 (1990).

\item{10.} T. Kaneko, T. Jones, and B.A. Joyce (unpublished). 

\item{11.} W.K. Liu, S.M. Mokler, N. Ohtani, J. Zhang, and B.A. Joyce, 
Appl. Phys.
Lett. {\bf 60}, 56 (1992).

\item{12.} M. Hiroi, K. Koyama, T. Tatsumi, and H. Hirayama, Appl. 
Phys. Lett.
{\bf 60}, 1723 (1992).

\item{13.} S.M. Mokler, W.K. Liu, N. Ohtani, and B.A. Joyce, Appl. 
Phys. Lett.
{\bf 60}, 2255 (1992).

\item{14.} F.J. Lamelas, P.H. Fuoss, P. Imperatori, D.W. Kisker, G.B. 
Stephenson,
and S. Brennan, Appl. Phys. Lett. {\bf 60}, 2610 (1992). 

\item{15.} D.W. Kisker, G.B. Stephenson, P.H. Fuoss, F.J. Lamelas, S. 
Brennan, and
P. Imperatori, J. Crystal Growth (in press). 

\item{16.} A. Robertson, Jr., T.H. Chiu, W.T. Tsang, and J.E. 
Cunningham, J. Appl.
Phys. {\bf 64}, 877 (1988).

\item{17.} S.M. Gates and S.K. Kulkarni, Appl. Phys. Lett. {\bf 60}, 
53 (1992).

\item{18.} W.K. Burton, N. Cabrera and F.C. Frank, Philos. Trans. R. 
Soc. London,
Sect. A {\bf 243}, 299 (1951).

\item{19.} H.J.W. Zandvliet, H.B. Elswijk, D. D.Dijkkamp, E.J. van 
Loenen, and
J. Dieleman, J. Appl. Phys. {\bf 70}, 2614 (1991). 

\item{20.} T. Shitara, J. Zhang, J.H. Neave, and B.A. Joyce, J. Appl. 
Phys. {\bf
71}, 4299 (1992).

\item{21.} Presumably, the mobile precursor in this case is not 
TMG, but dimethylgallium or monomethylgallium. See, e.g., Refs. 
7 and 8.

}

\vfill
\eject

\centerline{\bf Figure Captions}

\noindent
{\bf Figure 1.} Schematic illustration of the dominant kinetic 
processes included
in the model defined by equations (1)--(4). \bigskip

\noindent
{\bf Figure 2.} Kinetic phase diagram of the different growth 
regimes exhibited by
the model defined by equations (1)--(4) as a function of molecular 
deposition
flux (J) and
substrate temperature (T). Each phase boundary is the locus of 
points where the relevant ratio of length scales is unity. 

\bye